\documentclass[onecolumn,3p,a4paper,10pt]{elsarticle}
\usepackage{hyperref}
\usepackage{amssymb,natbib,amsmath,wasysym}
\usepackage{newtxtext,newtxmath}
\usepackage{caption}
\usepackage{subfigure}
\usepackage{url}
\usepackage{ulem}
\usepackage{tikz}
\usepackage{color}
\biboptions{numbers,sort&compress}
\usepackage{graphicx}
\newcommand{\mnras}{Monthly Notices of the Royal Astronomical Society}
\newcommand{\apj}{The Astrophysical Journal}
\newcommand{\apjs}{The Astrophysical Journal Supplement Series}
\newcommand{\aj}{Astronomical Journal}

\usepackage{orcidlink}

\begin{document}

\begin{frontmatter}

\title{ Modified-gradient methods for exact divergence-free in meshless magnetohydrodynamics}

\author[1]{Xiongbiao Tu}
\ead{hbnutu@yeah.net}
\author[2,5]{Qiao Wang \corref{cor1}\orcidlink{0000-0003-2153-7758}}
\ead{qwang@nao.cas.cn}
\author[3,6,5]{Liang Gao} 
\author[4,7]{Yifa Tang}

\address[1]{School of Mathematics and Statistics, Hubei Normal University, Huangshi 435002, China}
\address[2]{National Astronomical Observatories, Chinese Academy of Sciences, Beijing 100101, China}
\address[3]{Institute for Frontiers in Astronomy and Astrophysics, Beijing Normal University, Beijing 100875, China}
\address[4]{State Key Laboratory of Mathematical Sciences, Academy of Mathematics and Systems Science, Chinese Academy of Sciences, Beijing 100190, China}
\address[5]{School of Astronomy and Space Science, University of Chinese Academy of Sciences, Beijing, 100048, China}
\address[6]{School of Physics and Laboratory of Zhongyuan Light, Zhengzhou University, Zhengzhou 450001, China}
\address[7]{School of Mathematical Sciences, University of Chinese Academy of Sciences, Beijing 100049, China}

\cortext[cor1]{Corresponding author}

\begin{abstract}

We present a novel gradient regularization to completely eliminate the magnetic divergence error in meshless  magnetohydrodynamics (MHD), which offers a high spatial resolution and conservative advantage, due to its Lagrangian nature.  Comparing with the counterpart of constrained-gradient (CG) technique, we  reform $\nabla \cdot \mathbf{B}=0$ by  an implicit projection method to modify the magnetic-field gradients.

The accuracy of modified-gradient (MG) method  is verified and it achieves exact divergence-free results with round-off precision, by using  tests of shock tube, 2D and 3D vortex,  magneto-rotational instability, and especially, advection experiment, compared with CG method and the {\footnotesize GIZMO} code.  It leads to noticeable improvement in pattern, amplitude and numerical dissipation of divergence error of magnetic field.
\end{abstract}


\end{frontmatter}

\section{Introduction}

Ideal magnetohydrodynamics (MHD) coupling the Euler equations of an inviscid, non-resistive gas with Maxwell’s equations, well describes  high-temperature plasmas and astrophysical flows across disciplines\cite{Marinacci+2018,athena2020}.
Its formula can be expressed in the form of hyperbolic conservation laws. Numerical simulations practically employ its hyperbolic conservative form. 
\citet{Godunov+1972} found that the conservative equations could be symmetrizable to obtain {Galilean} invariant, by adding the constraint of $\nabla \cdot \mathbf{B} = 0$. {For the purpose of numerical research, one naturally look for discrete systems that} retain the characteristic attributes and internal symmetry of the original system as much as possible.

One of the essential challenges is to maintain a divergence-free of magnetic field in numerical MHD simulations. The early review paper \citep{Toth2000} gave a detailed comparison of seven preserving $\nabla \cdot \mathbf{B}=0$ schemes, including the eight-wave formulation, the projection scheme, and five different versions of the constrained transport (CT) type schemes. The popular MHD software package {\footnotesize ATHENA} and {\footnotesize ATHENA++}, based on adaptive mesh refinement algorithm, provides a excellent CT implementation \cite{Stone2008,athena2020}  on regular mesh. In addition, the solvers for MHD problem and the CT schemes maintaining the divergence-free constraint have achieved fruitful results \cite{Hawley1995,Rossmanith+2006, Kawai2013,LOPES2018293,Helzel+2011}. Some recent works \cite{Ding+2024,Li+2019,Li+2021} based on finite element methods have provided a new brunch for the MHD problems.

Unfortunately, it is very difficult to develop CT schemes in unstructured mesh, meshless method or moving  mesh  methods~\citep{Mocz+201401,Mocz+2016}, so that it leads to alternative divergence-cleaning schemes. The ``8-wave" cleaning scheme proposed by Powell et al. \cite{Powell1999} plays an important role in the stability of the divergence-cleaning schemes. The extra cleaning scheme was proposed by Dedner et al. \cite{DEDNER2002645} to effectively control the divergence error. Moving mesh methods with cleaning schemes for MHD have been developed and widely applied \cite{Duffell+2011,Pakmor+2011,Gaburov+2012,Pakmor+2013}. 
These schemes can control the divergence errors to a certain extent, but they alter the MHD equations and may make the numerical method non-conservative.

On the other hand,  the Lagrangian meshless algorithm is are developed and widely used due to their good conservation and automatic adaptive resolution. \citet{Lucy+1977} and \citet{Gingold+1977}  formulated the first meshless method named Smooth Particle Hydrodynamics (SPH), then it is developed by {B{\o}rve} et~al. \cite{Borve2001,Springel+2005}. \citet{Tricco+2012,Stasyszyn2013} explore the divergence-cleaning schemes for the in SPH scheme of MHD. Now cleaning terms become a powerful numerical method for the simulation of astrophysical problems. Further, the robust well-designed meshless methods for MHD have been studied  \citep{Gaburov+2011,Tricco+2013,Hopkins+2016a,Hopkins+2016b}, but they still require clean terms. So far, divergence free  ($\nabla \cdot \mathbf{B} = 0$) simulation technique remains a challenging problem on  meshless numerical MHD.

If the divergence error is not effectively eliminated, the direction of the Lorentz force is not perpendicular to the direction of magnetic field in solving the MHD problem, resulting in numerical dissipation and non-physical numerical solutions \cite{Brackbill+1980,Helzel+2011}.
{Its} impact on the simulation will  beyond a loss of accuracy, it will also compromise numerical stability.
Theoretically, the magnetic field can be derived from electro-magnetic potential and it always keep divergence-free.
In this branch, the feasibility of potential method to evolve magnetic field has been explored, such as Euler potential (\citet{Brandenburg+2010}), vector potential (\citet{Price+2010}, \citet{Stasyszyn2015}), potential methods for Maxwell equations (\citet{Xiao+2013}, \citet{Qin+2016}), as well as the recent work (\citet{Tu+2022}) providing a vector potential method for numerical MHD problems based on the robust meshless method (\citet{Hopkins+2016a}).

In this paper, we introduce to a Lagrangian Godunov-type meshless modified-gradient (MG) method for MHD to {eliminate} the divergence error. It is an modification for the robust meshless constrained-gradient (CG) code of \citet{Hopkins+2016b}. The outline of this paper is as follows. In Section 2, we briefly present the MHD equations and the  Lagrangian Godunov-type meshless method. The numerical implementation of the MG method is described in Section 3. Some technical details of the MG method are described in Section 4. Several  numerical experiments are present in Section 5. Finally, we summarize the results in Section 6.

\section{Numerical method}
\subsection{The magnetohydrodynamic equations}
\label{sect:2}

The ideal MHD equations without viscosity and resistivity governs the physics of the ideal magnetized plasma fluid coupling with the magnetic field. They contain continuity, momentum, energy equations, and induction equations and  works for variant physical conditions and the hyperbolic conservation version of the equations is suitable for numerical simulation. Conventionally, the notation can be written as 
\begin{equation}
  \label{eq:cons}
L_{v}(\textbf{U}) + \nabla\cdot\mathbf{F} = 0,
\end{equation}
\begin{equation}
  \label{eq:Bres}
\nabla \cdot \mathbf{B} = 0,
\end{equation}
where $L_{v}(\textbf{U}) = \frac{\partial\textbf{U}}{\partial t} + \nabla\cdot(\mathbf{v}_{\rm f}\otimes\mathbf{U})$. The vector of conserved variables $\mathbf{U}$ and flux $\mathbf{F} = F - \mathbf{v}_{\rm f}\otimes\mathbf{U}$ reads
\begin{equation} \label{eq:2.1.2}
\textbf{U}=
\left[\begin{array}{c}
   \rho  \\
   \rho \mathbf{v} \\
   \rho e \\
   \mathbf{B}
  \end{array}\right],~ \text{and}~
F =
\left[\begin{array}{c}
   \rho \mathbf{v} \\
   \rho \mathbf{v}\otimes \mathbf{v} + P_{T}\mathcal{I} - \mathbf{B}\otimes\mathbf{B} \\
   (\rho e + P_{T})\mathbf{v} - (\mathbf{v}\cdot \mathbf{B})\mathbf{B} \\
   \mathbf{v}\otimes \mathbf{B} - \mathbf{B}\otimes \mathbf{v}\
   \end{array}\right],
\end{equation}
respectively, where $\rho$ is the mass density, $e = u + |\mathbf{B}|^2 /(2\rho) + |\mathbf{v}|^2 /2$ is the total specific energy, $u$ is the internal energy, $P_T = {P_{\rm gas}} + |\mathbf{B}|^2 /2$ is the total pressure including thermal and magnetic pressures.
The ideal magnetized plasma fluid obeys the equation of state ${P_{\rm gas}}=(\gamma - 1)\rho u$, where $\gamma$ is a constant  polytropic coefficient.
The notation $\otimes$ is a outer product. The exact solution of the ideal MHD equations (\ref{eq:cons}) obeys the divergence-free condition (\ref{eq:Bres}) . Because there are no magnetic monopoles and it Mathematically keeps the equations symmetric and Galilean invariant.

\subsection{Conservative meshless method}
Following the works of \citet{Lanson+2008a} and \citet{Gaburov+2011}, we discretize MHD equations (\ref{eq:cons}) into a weighted meshless scheme with a continuous symmetric kernel function $W(\mathbf{x}-\mathbf{x}_{i}, h(\mathbf{x}))$, which is  normalized $\int_{\mathbb{R}^d}W(\mathbf{x}-\mathbf{x}^{\prime}, h(\mathbf{x}^{\prime})) d\mathbf{x}= 1$ at compact support. The fraction function $\psi_{i}(\mathbf{x})$ around particle $i$ is defined by
\begin{equation} \label{eq:2.2.1}
\psi_i(\mathbf{x}) = \frac{1}{\omega(\mathbf{x})}W(\mathbf{x} - \mathbf{x}_i,h(\mathbf{x})),
\end{equation}
where $\mathbf{x}_i$ is the position of particle $i$, $\omega(\mathbf{x}) = \sum_{j\in P} W(\mathbf{x}-\mathbf{x}_j,h(\mathbf{x})) $ is the normalization with the kernel size $h(\mathbf{x})$ of function $W$, $\sum_{i\in P}\psi_i(\mathbf{x}) \equiv 1$. The notation of $P$ is the particle set over the domain under consideration. The meshless derivatives of a scalar function with respect to $x^{\alpha}$, $\alpha \in \{1,2,3\}$ can be estimated as follows
\begin{equation} \label{eq:2.2.2}
\begin{split} 
(\nabla f)_i^{\alpha} &= \sum_{j}(f_j-f_i )\textbf{T}^{\alpha\beta}_i(\mathbf{x}_{j}^{\beta}-\mathbf{x}_i^{\beta}) \psi_{j}(\mathbf{x}_i)\\
&= \sum_{j}(f_j-f_i)\tilde{\psi}_{j}^{\alpha}(\mathbf{x}_i),
\end{split}
\end{equation}
where $j$ refers to all neighbours of particle $i$, we use an Einstein summation convention over the indice $\beta$, $\tilde{\psi}_{j}^{\alpha}(\mathbf{x}_i) = \textbf{T}^{\alpha\beta}_i(\mathbf{x}_{j}^{\beta}-\mathbf{x}_i^{\beta}) \psi_{j}(\mathbf{x}_i)$. The components of a $3\times3$ matrix $\textbf{E}^{\alpha\beta}_i = \sum_{j\in P}(\mathbf{x}_{j}-\mathbf{x}_i)^{\alpha}(\mathbf{x}_{j}-\mathbf{x}_i)^{\beta}\psi_j(\mathbf{x}_i),$ $\alpha,\beta=1,2,3$ and $\textbf{T}_i = \textbf{E}_i^{-1}$. This estimator has been widely used in various works~\citep{Onate+1996, Lanson+2008a, Lanson+2008b, Luo+2008, Gaburov+2011, Hopkins+2015}.

We solve the equations at a moving frame ${\mathbf v}_{\rm f}$ so that the transport operator is  $L_{v}(\mathbf{U}) = {\partial \mathbf{U}}/{\partial t} + \nabla \cdot( {\mathbf v}_{\rm f}\otimes \mathbf{U})$ . Thus, the weak solution of the conservative equation (\ref{eq:cons}) can be derived . The conservative equation is multiplied by a continuously differentiable test function $\varphi$,
$\varphi \in C_{0}^{1}(\mathbb{R}^d\times\mathbb{R}^{+})$, and integrated over the compact support
\begin{equation} 
\int_{\mathbb{R}^d\times\mathbb{R}^{+}}(L_{v}(\mathbf{U})\varphi + \nabla\cdot\mathbf{F}(\textbf{U}) \varphi ) dxdt = 0.
\end{equation}
Considering an integration by parts and Gauss Theorem, the following integral form is derived
\begin{equation} 
\int_{\mathbb{R}^d\times\mathbb{R}^{+}}(\textbf{U}L_{v}^{\ast}(\varphi) + \mathbf{F}(\textbf{U})\cdot\nabla \varphi ) dxdt = 0,
\end{equation}
where $L_{v}^{\ast}(\varphi) = \frac{\partial \varphi}{\partial t} + {\mathbf v}_{\rm f}\cdot\nabla\varphi$. 
We transfer the spatial integral and meshless derivatives of $\varphi$ to the integral function and obtain
\begin{equation}
\int_{\mathbb{R}^{+}} \sum_{i\in P}\left[V_{i}\textbf{U}_{i}\frac{d\varphi_{i}}{dt} + \sum_{j}V_{i}\mathbf{F}^{\alpha}(\textbf{U}_{i})(\varphi_{j}-\varphi_{i})\tilde{\psi}_{j}^{\alpha}(\mathbf{x}_i) \right] dt = 0,
\end{equation}
where $j$ refers to all interacting neighbours $j$ of $i$, $V_i$ is the ``effective volume'' of particle $i$, which is estimated by ${1}/{\omega(x_i)}$, keeping the same as the previews works of \citet{Hopkins+2015} and \citet{Tu+2022}. 

Again, by using integration by parts, straightforward calculation and reordering, we derived the following integral form of
\begin{equation} 
\int_{\mathbb{R}^{+}} \sum_{i\in P}\left[\frac{d }{dt}(V_{i}\textbf{U}_{i}) + \sum_{j} \left( V_{i}\mathbf{F}^{\alpha}(\textbf{U}_{i})\tilde{\psi}_{j}^{\alpha}(\mathbf{x}_i) - V_{j}\mathbf{F}^{\alpha}  ( \textbf{U}_{j})\tilde{\psi}_{i}^{\alpha}(\mathbf{x}_j) \right)\right]\varphi_{i} dt = 0.
\end{equation}
For any test function $\varphi$, the above equation holds. Therefore, the expression inside the parenthesis must be zero, i.e.,
\begin{equation} 
\label{eq:2.2.7}
\frac{d }{dt}(V_{i}\textbf{U}_{i}) + \sum_{j} \left( V_{i}\mathbf{F}^{\alpha}(\textbf{U}_{i})\tilde{\psi}_{j}^{\alpha}(\mathbf{x}_i) - V_{j}\mathbf{F}^{\alpha}(\textbf{U}_{j})\tilde{\psi}_{i}^{\alpha}(\mathbf{x}_j) \right) = 0.
\end{equation}

For interactive pair of particles $i$ and $j$, we enforce the fluxes $\tilde{\mathbf{F}}_{ij}$ to be calculated on the virtual interface $\mathbf{x}_{ij}=\frac{1}{2}(\mathbf{x}_i+\mathbf{x}_j)$ by an appropriate flux solver, and  take the place of $\mathbf{F}(\mathbf{U}_{i})$ and $\mathbf{F}(\mathbf{U}_{j})$ with $\tilde{\mathbf{F}}_{ij}$. 
We define normal vector of,  $\mathbf{\hat{A}}_{ij}^{\alpha} = V_{i}\tilde{\psi}_{j}^{\alpha} (\mathbf{x}_{i})-V_{j} \tilde{\psi}_{i}^{\alpha}(\mathbf{x}_{j}), \alpha=1,2,3$,  the virtual interface  and  its area  is $|\mathbf{\hat{A}}_{ij}|$. The discrete equation (\ref{eq:2.2.7}) can be expressed as follows
\begin{equation}
  \label{eq:2.2.8}
\frac{d }{dt}(V_{i}\mathbf{U}_{i}) + \sum_{j}\tilde{\mathbf{F}}_{ij}\cdot\mathbf{\hat{A}}_{ij}  = 0,
\end{equation}
where $j$ refers to  interactive neighbours particle about particle $i$.

For each particle, we set the quantities (“conservative variables”): the mass $m_i$, momentum $\mathbf{P}_i$, total energy $E_i$, and weighted magnetic field ${\mathcal{B}}_i$, given by
\begin{equation}
  \label{eq:2.2.9}
Q_i = 
\left(\begin{array}{c}
   m_i  \\
   \mathbf{P}_i \\
   E_i \\
   {\mathcal{B}}_i
  \end{array}\right)
= 
V_i\mathbf{U}_{i}.
\end{equation}
Equation (\ref{eq:2.2.8}) is discretized as follows:
\begin{equation}
  \label{eq:2.2.10}
Q^{(n+1)}_i = Q^{(n)}_i + \Delta t \sum_{j}\hat{\mathbf{F}}_{ij}^{(n+1)},
\end{equation}
where $\hat{\mathbf{F}}_{ij}=-\tilde{\mathbf{F}}_{ij}\cdot\mathbf{\hat{A}}_{ij}$ and $\hat{\mathbf{F}}_{ij}^{(n+1)}$ is an appropriate estimate of the numerical flux at the end of a timestep.
Due to its antisymmetric, $\hat{\mathbf{F}}_{ij}^{(n+1)} = -\hat{\mathbf{F}}_{ji}^{(n+1)}$, our meshless method still holds conservative properties.

\section{Numerical implementation}
\subsection{Kernel sizes and particle volumes}

In the meshless scheme, the kernel size is determined by a fixed neighboring particles number in the adaptive support  region.  In detail, we follow \citet{Gaburov+2011,Hopkins+2015} as the equation of 
\begin{equation}
  \label{eq:NGB}
N_{\textup{NGB}} = C_{\nu}h^{\nu}(\mathbf{x}_i)\sum_{j\in P}W(\mathbf{x}_j-\mathbf{x}_i,h(\mathbf{x}_i)),
\end{equation}
where  $C_{\nu} = \pi$ and  $N_{\textup{NGB}} = 20$ for two dimension case, $\nu = 2$ and $C_{\nu} = 4\pi/3$ and $N_{\textup{NGB}} = 32$ for three dimension. Practically, the kernel size $h(\mathbf{x}_i)$ needs to be iteratively calculated by formulation (\ref{eq:NGB}). Then, the “effective volume” of particles can be defined by
\begin{equation}
  \label{eq:3.1.2}
V_i = \omega^{-1}(\mathbf{x}_i) = \left\{ \sum_{j\in P} W \left(\mathbf{x}_i-\mathbf{x}_j,h(\mathbf{x}_i) \right) \right\}^{-1}.
\end{equation}
Moreover, the density is estimated by  $\rho_i= m_i/V_i$.

\subsection{Timestep}
We use a uniform timestep. The timestep is set according to a standard local Courant-Friedrichs-Levy ({\rm CFL}) timestep criterion, as follows
\begin{equation}
  \label{eq:3.2.1}
\Delta t < \min_{i,j}\left\{ C_{\rm CFL}\frac{2R_i}{2c_{f,i} + |{\mathbf v}_i - {\mathbf v}_j|}\right\},
\end{equation}
where $R_i$ is the effective radius of particle $i$, $j$ refers to all interacting neighbours $j$ of $i$, $i\in P$. For the stable execution of MHD simulation, we set $C_{\rm CFL} = 0.4$. The fast magnetosonic speed $c_{f,i}$ in denominator is calculated by 
\begin{equation} 
\label{eq:3.2.2}
c_{f}=  \left\{  \frac{a^2}{2} + \frac{|\mathbf B|^2}{2\rho} + \sqrt{\left( \frac{a^2}{2}   + \frac{|\mathbf B|^2}{2\rho} \right)^2 - \frac{ a^2 B_{m}^{2}}{\rho}}  \right\}^{1/2} ,
\end{equation}
where $a=\sqrt{{\gamma {P_{\rm gas}}}/{\rho}}$ is the sound speed, $B_{m}$ is the projection of  ${\mathbf B}$ along the pair-wise direction.

\subsection{The discrete gradients and slope limiter}
We estimate the gradient values of the primitive variables $(\rho, {\mathbf v}, {P_{\rm gas}}, {\mathbf B})$ of each particle by equation (\ref{eq:2.2.2}). Specifically, the gradient in particle $i$ of a primitive variable $f$ is determined by
\begin{equation} \label{eq:3.3.1}
(\nabla f)_i = \sum_{j}(f_j-f_i) \textbf{T}_i (x_{j}-x_i) \psi_{j}(x_i),\\
\end{equation}
where the subscript of $j$ refers to interacting neighbor particle about particle $i$. {The} $3\times3$ matrix $\textbf{T}_i  = \textbf{E}_i^{-1}$ and the components of matrix $\textbf{E}_i^{\alpha\beta} = \sum_{j\in P}(x_{j}-x_i)^{\alpha}(x_{j}-x_i)^{\beta}\psi_j(x_i), ~ \{\alpha,\beta = 1,2,3\}$.

Discontinuities, such as shocks, causing severe oscillations in the numerical solution and even affecting the stability of evolution. We also introduce a slope limiter of the  primitive variables.  
We present the mathematical form, as follow
\begin{equation} \label{eq:3.3.2}
(\nabla f)_{{\rm lim},i} = \min(1,t_i)(\nabla f)_{i}, \\
\end{equation}
where $t_i$ depends on the maximum value $f^{{\rm max}}_i$ and minimum value $f^{{\rm min}}_i$ of $f$ on all particles in the support region. The specific form of $t_i$ is as follows
\begin{equation} \label{eq:3.3.2}
t_i = \min \left(\frac{f^{{\rm max}}_i-f_i}{TV^{{\rm max}}_i},\frac{f^{{\rm min}}_i-f_i}{TV^{{\rm min}}_i} \right), 
\end{equation}
where $TV^{{\rm max}}_i=\max_j\{(\nabla f)_{i}(\textbf{x}_{ij}-\textbf{x}_i)\}$, $TV^{{\rm min}}_i=\min_j\{(\nabla f)_{i}(\textbf{x}_{ij}-\textbf{x}_i)\}$ and $\mathbf{x}_{ij}=\frac{1}{2}(\mathbf{x}_i+\mathbf{x}_j)$  at virtual interfaces.

\subsection{The numerical flux}
The numerical flux for the mass $m_i$, momentum $\mathbf{P}_i$, total energy $E_i$, and weighted magnetic field ${\mathcal{B}}_i$ is calculated in the rest frame of each of the virtual interface. We first move the particles over the time interval $\Delta t$ based on their velocities and then predict fluid quantities $m, {\mathbf v}, u, {\mathcal{B}}$ at {$t_n + \Delta t$}. This can be calculated using the following formula
\begin{equation} \label{eq:3.4.1}
\hat{f}_i = f_i + \Delta t{\frac{d f_{i}^{(n)}}{dt}},
\end{equation}
we will explain ${d f_{i}^{(n)}/{dt}}$ in the end of this subsection and $f_i$ represents the fluid quantities $m_i, {\mathbf v}_i, u_i, {\mathcal{B}}_i$.

We convert the predicted fluid quantities to primitive variables ${\mathbf W}^{\top} = (\rho, {\mathbf v}, {P_{\rm gas}}, {\mathbf B})$. Next, we calculate the discrete gradients by the formula (\ref{eq:3.3.1}) to the primitive variables, and then extrapolate the primitive variables to the virtual interface for flux calculation at the end of the timestep. We linearly reconstruct the primitive variables on both sides of the virtual interface $\mathbf{x}_{ij}=\frac{1}{2}(\mathbf{x}_i+\mathbf{x}_j)$, 
\begin{equation} \label{eq:3.4.2}
\begin{split}
{\mathbf W}_i^{\prime} &= {\mathbf W}_i + (\nabla \mathbf W)_{{\rm lim},i}(\mathbf{x}_{ij} - \mathbf{x}_i),\\
{\mathbf W}_j^{\prime} &= {\mathbf W}_j + (\nabla \mathbf W)_{{\rm lim},j}(\mathbf{x}_{ij} - \mathbf{x}_j).
\end{split}
\end{equation}
For neighbouring particles $i$ and $j$, we estimate the virtual interface velocity $\mathbf{v}_{ij} = \frac{1}{2}({\mathbf v}_i + {\mathbf v}_j)$. We then change the reconstructed values of primitive variables from the lab frame to the rest frame of the virtual interface
\begin{equation} \label{eq:3.4.3}
\begin{split}
{\mathbf W}_i^{\prime\prime} &= (\rho_i^{\prime}, {\mathbf v}_i^{\prime} - \mathbf{v}_{ij}, ({P_{\rm gas}})_i^{\prime}, {\mathbf B}_i^{\prime}),\\
{\mathbf W}_j^{\prime\prime} &= (\rho_j^{\prime}, {\mathbf v}_j^{\prime} - \mathbf{v}_{ij}, ({P_{\rm gas}})_j^{\prime}, {\mathbf B}_j^{\prime}).
\end{split}
\end{equation}

Following \citet{Springel+2010} and \citet{Mocz+201401}, we compute the flux {at} a new cartesian coordinate frame $(\hat{e}_1, \hat{e}_2, \hat{e}_3)$, whose $\hat{e}_1$ is parallel to the normal vector $\mathbf{\hat{A}}_{ij}$ of virtual interface. Then the state is rotated into the new coordinate frame 
\begin{equation} \label{eq:3.4.4}
{\mathbf W}_k^{\prime\prime\prime} = 
\left(\begin{array}{cccc}
   1 & 0 & 0 & 0  \\
   0 & \Lambda & 0 & 0 \\
   0 & 0 & 1 & 0 \\
   0 & 0 & 0 &\Lambda  \\
  \end{array}\right){\mathbf W}_k^{\prime\prime},
\end{equation}
where $\Lambda = (\hat{e}_1, \hat{e}_2, \hat{e}_3)^{\top}$ is a $3\times3$ matrix, $k=i,j$. In general, in the new coordinate frame, the components $B_{x,i}^{\prime\prime\prime}$ and $B_{x,j}^{\prime\prime\prime}$ of magnetic fields in the $\hat{e}_1$ direction are discontinuous, we enforce the mean value {of} this quantity  across the virtual face
\begin{equation} \label{eq:3.4.5}
{\bar{B}}_{x,ij}^{\prime\prime\prime} = \frac{1}{2}(B_{x,i}^{\prime\prime\prime} + B_{x,j}^{\prime\prime\prime}).
\end{equation}
We convert the primitive variables ${\mathbf W}_i^{\prime\prime\prime}$, ${\mathbf W}_j^{\prime\prime\prime}$ to the conservative variables ${\mathbf U}_i^{\prime\prime\prime}$, ${\mathbf U}_j^{\prime\prime\prime}$.
In the new coordinate frame, we approximately solve the Riemann problem using a 1-D Harten-Lax-van Leer Discontinuities (HLLD) Riemann solver (\citet{Miyoshi+2005}) to calculate the flux 
\begin{equation} \label{eq:3.4.6}
{\mathbf F}_{ij}^{\prime\prime\prime} = F_{\rm HLLD}({\mathbf U}_i^{\prime\prime\prime},{\mathbf U}_j^{\prime\prime\prime}).
\end{equation}
We then rotate  the fluxes ${\mathbf F}_{ij}^{\prime\prime\prime}$ from the new coordinate frame to the rest frame 
\begin{equation} \label{eq:3.4.7}
{\mathbf F}_{ij}^{\prime\prime} = 
\left(\begin{array}{cccc}
   1 & 0 & 0 & 0  \\
   0 & \Lambda^{\top} & 0 & 0 \\
   0 & 0 & 1 & 0 \\
   0 & 0 & 0 &\Lambda^{\top} \\
  \end{array}\right){\mathbf F}_{ij}^{\prime\prime\prime}.
\end{equation}
We rewrite ${\mathbf F}_{ij}^{\prime\prime}$ as a vector of  flux $(F_1,F_2,F_3,F_4,F_5,F_6,F_7,F_8)^{\top}$ so that {we} can convert the flux ${\mathbf F}_{ij}^{\prime\prime}$ from the rest frame to the lab frame by adding an additional correction term due to the movement of the virtual interface, given by Pakmor et al. \cite{Pakmor+2011}
\begin{equation} \label{eq:3.4.8}
\hat{\mathbf{F}}_{ij}^{(n+1)}={\mathbf F}_{ij}^{\prime\prime} +
\left(\begin{array}{c}
   0  \\
   v_{x,ij}F_1\\
   v_{y,ij}F_1\\
   v_{z,ij}F_1\\
   v_{x,ij}F_2 + v_{y,ij}F_3 + v_{z,ij}F_4 + \frac{1}{2}F_1\mathbf{v}_{ij}^2\\
   -v_{x,ij}{\bar{B}}_{x,ij}^{\prime\prime\prime} \\
   -v_{y,ij}{\bar{B}}_{x,ij}^{\prime\prime\prime} \\
   -v_{z,ij}{\bar{B}}_{x,ij}^{\prime\prime\prime} \\
  \end{array}\right).
\end{equation}
Now, we have obtained the numerical flux $\hat{\mathbf{F}}_{ij}^{(n+1)}$ required to update $Q^{(n)}_i$ in {the discreted equation (\ref{eq:2.2.10}).}

{Finally, we return to the discussion on the predicting fluid quantities mentioned, and provide an estimate for $d f_{i}^{(n)}/{dt}$ in Eq. (\ref{eq:3.4.1}). }We rewrite $\hat{\mathbf{F}}_{i}^{(n+1)} = \sum_{j}\hat{\mathbf{F}}_{ij}^{(n+1)}$ and  $\hat{\mathbf{F}}_{i}^{(n+1)}$ as $(\hat{F}_1,\hat{F}_2,\hat{F}_3, \hat{F}_4, \hat{F}_5, \hat{F}_6, \hat{F}_7, \hat{F}_8)^{\top}$. According to Eq. (\ref{eq:2.2.8}), we can obtain $dm_{i}^{(n+1)}/dt = \hat{F}_1$, $d\mathbf{P}_i^{(n+1)}/dt=(\hat{F}_2,\hat{F}_3,\hat{F}_4)^{\top}$, $dE_{i}^{(n+1)}/dt=\hat{F}_5$, $d{\mathcal{B}}_i^{(n+1)}/dt=(\hat{F}_6,\hat{F}_7,\hat{F}_8)^{\top}$. The estimations for {$d\mathbf{v}_{i}^{(n+1)}/dt$, $du_i^{(n+1)}/dt$} in Eq. (\ref{eq:3.4.1}) are given by the following equations
\begin{equation}
  \label{eq:3.4.9}
\begin{aligned}
\frac{d\mathbf{v}_{i}}{dt} &= \frac{1}{m_i}\left(\frac{d\mathbf{P}_i}{dt} - \mathbf{v}_{i}\frac{dm_{i}}{dt} \right), \\
\frac{du_i}{dt} &= \frac{1}{m_i} \left(\frac{dE_i}{dt} - \mathbf{v}_{i}\frac{d\mathbf{P}_i}{dt} - \mathbf{B}_{i}\frac{d{\mathcal{B}}_{i}}{dt} + \frac{1}{2}\mathbf{v}_{i}^2\frac{dm_{i}}{dt}  + \frac{1}{2}\mathbf{B}_{i}^2\frac{dV_{i}}{dt} -u_i\frac{dm_{i}}{dt} \right).\\
\end{aligned}
\end{equation}

{
\section{Modified gradient of magnetic field}

The above processing in the previous section cannot fully eliminate the  divergence of magnetic field. We found that the  reconstructed value of magnetic field needs to be modified to vanish magnetic flux in a closed surface. According to Gauss's law, the divergence of a magnetic field can be estimated as follows
\begin{equation} \label{eq:4.2}
\nabla\cdot \mathbf{B}_i = \frac{1}{V_i}\sum_j\Phi_{ij},
\end{equation}
where $j$ refers to interacting neighbours $j$ surrounding particle $i$ and $\Phi_{ij}$ is the normal magnetic flux from particle $i$ through the virtual surface between particle $i$ and particle $j$. In detail, $\Phi_{ij}$ is estimated by
\begin{equation} \label{eq:4.3}
\Phi_{ij}= {\bar{B}}_{x,ij}^{\prime\prime\prime} |\mathbf{\hat{A}}_{ij}| =\frac{1}{2}(B_{x,i}^{\prime\prime\prime} + B_{x,j}^{\prime\prime\prime})|\mathbf{\hat{A}}_{ij}| =  \frac{1}{2}(\mathbf{B}_i^{\prime} + \mathbf{B}_j^{\prime})\mathbf{\hat{A}}_{ij},
\end{equation}
where $\mathbf{B}_i^{\prime}$, $\mathbf{B}_j^{\prime}$ is the reconstructed values of magnetic field on both sides of the virtual interface, by Eq.~\ref{eq:3.4.2}. Moreover, the divergence  $\nabla\cdot \mathbf{B}_i$ reads
\begin{equation} \label{eq:4.4}
\nabla\cdot \mathbf{B}_i = \frac{1}{V_i}\sum_j\frac{1}{2}(\mathbf{B}_i^{\prime} + \mathbf{B}_j^{\prime})\mathbf{\hat{A}}_{ij}.
\end{equation}
This form is the conventional definition in most of finite volume method.

Note that Powell's scheme \citep{Powell1999} have to be employed a cleaning schemes to be compatible with magnetic field divergence, the formula symmetry and numerical stability. For a certain particle $i$, the source vector of $\mathbf{S}_{\text{Powell}}$ reads
\begin{equation}
\mathbf{S}_{\text{Powell}}=-(V\nabla\cdot \mathbf{B})_i\left[\begin{array}{c}
   0  \\
   \mathbf{B}_i \\
   \mathbf{v}_i\cdot\mathbf{B}_i\\
   \mathbf{v}_i 
\end{array}\right].
\end{equation}
In general, the Powell's {scheme} itself does not yield numerical convergence results, and additional source terms are often required to suppress magnetic field divergence error, such as the Dedner's term \citep{DEDNER2002645}. 
Anyway, the extra cleaning term cause non-conservative numerical methods.

In our method, we rewrite Eq.~\ref{eq:4.4} as  
\begin{equation} \label{eq:4.5}
\nabla\cdot \mathbf{B}_i = \frac{1}{2V_i} \sum_j \biggl[{\mathbf B}_i + (\nabla\otimes {\mathbf B})_i \cdot(\mathbf{x}_{ij}-\mathbf{x}_i) + {\mathbf B}_{j} + (\nabla\otimes {\mathbf B})_j \cdot(\mathbf{x}_{ij}-\mathbf{x}_j)\biggr]\mathbf{\hat{A}}_{ij},
\end{equation}
where $\mathbf{x}_{ij}=\frac{1}{2}(\mathbf{x}_i+\mathbf{x}_j)$ and discrete gradients $\nabla\otimes {\mathbf B}$ of  ${\mathbf B}$-field. Further, we use $(\nabla\otimes {\mathbf B})_i + c_i Q_{ij}$ and $(\nabla\otimes {\mathbf B})_j + c_j Q_{ji}$  to replace $(\nabla\otimes {\mathbf B})_i$ and $(\nabla\otimes {\mathbf B})_j$ in Eq.~\ref{eq:4.5}. It will introduce a new degree of freedom for each particle $i$. Define $Q_{ij} = -(\mathbf{x}_{ij}-\mathbf{x}_i)\otimes \mathbf{\hat{A}}_{ij}$. It is obviously symmetric $Q_{ij}=Q_{ji}$ due to $\mathbf{\hat{A}}_{ij} = -\mathbf{\hat{A}}_{ji}$. In our method, we constrain Eq.~\ref{eq:4.5} vanishing, then obtain the following equation
\begin{equation} \label{eq:4.6}
\begin{split}
\sum_j \bigg[c_i Q_{ij}:Q_{ij} - c_j Q_{ji}:Q_{ji}\bigg] &= \sum_j\bigg[{\mathbf B}_i + (\nabla\otimes {\mathbf B})_i \cdot(\textbf{x}_{ij}-\textbf{x}_i) + {\mathbf B}_{j} + (\nabla\otimes {\mathbf B})_j \cdot(\textbf{x}_{ij}-\textbf{x}_j)\bigg]\mathbf{\hat{A}}_{ij}, \\
\end{split}
\end{equation}
where  the notation of `$:$' indicates that Element-wise multiply the corresponding components of the two $3\times 3$ tensor matrices, then sum the nine products to yield a scalar. The vector $S_i$ takes place of the right-hand part of Eq.~\ref{eq:4.6}, i.e., 
\begin{equation} \label{eq:4.7}
\begin{split}
\sum_j \bigg[c_i Q_{ij}:Q_{ij} - c_j Q_{ji}:Q_{ji}\bigg] = S_i. ~~ \{ i\in P \}
\end{split}
\end{equation}

Thus, we now obtain a symmetric sparse system of linear equations consisting of $N$ equations ($N$ is the total particle number of the set $P$):
\begin{equation} \label{eq:4.8}
 {\mathcal R} X = {\mathbf b}, 
\end{equation}
where the $i$-th component of ${\mathbf b}$ is $S_i$, the components of the coefficient matrix $ { {\mathcal R} }(i,i) = \sum_j Q_{ij}:Q_{ij}$, and
\begin{equation} 
{{\mathcal R}}  (i,j) = { {\mathcal R}} (j,i) = 
\left\{\begin{array}{c}
\begin{split}
   -Q_{ij}:Q_{ij},~~~&\text{interactive pair of particle (i,j) ,}\\
   0, ~~~~~~~~~~~~~~~~&\text{otherwise.}
\end{split}
  \end{array}\right.
\end{equation}

Note that the symmetry of the coefficient matrix in equation (\ref{eq:4.8}) is crucial for accurately solving and achievemnet of $\nabla\cdot \mathbf{B}_i = 0$. Practically, 
we use the parallel sparse solver of Intel Math Kernel Library (MKL) PARDISO to deal with the symmetric sparse linear equations (Eq.~\ref{eq:4.8}) .

Given the pair of interactive particles $i$ and $j$, and we can compute modified gradients $(\nabla\otimes {\mathbf B})_{i,\rm MG}$ and $(\nabla\otimes {\mathbf B})_{j,\rm MG}$, as follows
\begin{equation} \label{eq:4.9}
\begin{split}
(\nabla\otimes {\mathbf B})_{i,\rm MG} &= (\nabla\otimes {\mathbf B})_i + c_iQ_{ij},\\
(\nabla\otimes {\mathbf B})_{j,\rm MG} &= (\nabla\otimes {\mathbf B})_j + c_jQ_{ji}.
\end{split}
\end{equation}
We use $(\nabla\otimes {\mathbf B})_{i,\rm MG}$ and $(\nabla\otimes {\mathbf B})_{j,\rm MG}$ to update $(\nabla\otimes {\mathbf B})_i$ and $(\nabla\otimes {\mathbf B})_j$ in the above Eq.~\ref{eq:3.4.2}.
The updated  values ${\mathbf B}_{i,\rm MG}^{\prime}$ and ${\mathbf B}_{j,\rm MG}^{\prime}$ of magnetic field in the virtual interface are linearly reconstructed by the form of 
\begin{equation} \label{eq:4.10}
\begin{split}
{\mathbf B}_{i,\rm MG}^{\prime} &= {\mathbf B}_i + (\nabla\otimes {\mathbf B})_{i,\rm MG} \cdot(\mathbf{x}_{ij}-\mathbf{x}_i),\\
{\mathbf B}_{j,\rm MG}^{\prime} &= {\mathbf B}_j + (\nabla\otimes {\mathbf B})_{j,\rm MG} \cdot(\mathbf{x}_{ij}-\mathbf{x}_j).
\end{split}
\end{equation}
Subsequently, the divergence of magnetic field is expressed by
\begin{equation} \label{eq:4.11}
\nabla\cdot \mathbf{B}_{i,\rm MG} = \frac{1}{V_i}\sum_j\frac{1}{2}(\mathbf{B}_{i,\rm MG}^{\prime} + \mathbf{B}_{j,\rm MG}^{\prime})\mathbf{\hat{A}}_{ij}.
\end{equation}

Now, we present $\nabla\cdot \mathbf{B}_{i,\rm MG} = 0$ for any particle $i$ , in our modified gradient method. \\ [2mm]

\begin{tikzpicture}
\fill (0.0,0.0) circle (0.1pt);

\fill (2.0,0.0) circle (2.0pt);
\fill (4.0,0.0) circle (2.0pt);
\fill (6.0,0.0) circle (2.0pt);

\draw[very thick,->] (4.0,0.0) -- (3.2,1.5);
\draw[very thick,->] (4.0,0.0) -- (4.7,1.5);
\draw[very thick,->] (4.0,0.0) -- (5.7,1.);
\draw[thick,dashed] (2.0,0.0) -- (6,0.0);

\node at (2,-0.3) {$\mathbf{x}_i$};
\node at (4,-0.3) {$\mathbf{x}_{ij}$};
\node at (6,-0.3) {$\mathbf{x}_j$};
\node at (5.7,0.6) {$B_{x,i}^{\prime\prime\prime}$};
\node at (4.2,1.4) {$B_{y,i}^{\prime\prime\prime}$};
\node at (2.8,1.2) {$B_{z,i}^{\prime\prime\prime}$};
\node at (1.77,1.5) {${\mathbf B}_{i}^{\prime}:$};

\fill (9.0,0.0) circle (2.0pt);
\fill (11.0,0.0) circle (2.0pt);
\fill (13.0,0.0) circle (2.0pt);

\draw[very thick,->] (11.0,0.0) -- (10.2,1.5);
\draw[very thick,->] (11.0,0.0) -- (11.7,1.5);
\draw[very thick ,red,->] (11.0,0.0) -- (13.04,1.2);
\draw[very thick, ->] (11.0,0.0) -- (12.7,1.);
\draw[thick,dashed] (9.0,0.0) -- (13,0.0);

\node at (9,-0.3) {$\mathbf{x}_i$};
\node at (11,-0.3) {$\mathbf{x}_{ij}$};
\node at (13,-0.3) {$\mathbf{x}_j$};
\node at (12.6,0.5) {$B_{x,i}^{\prime\prime\prime}$};
\node[red] at (13.5,0.8) {$B_{x,i,\rm MG}^{\prime\prime\prime}$};
\node at (11.,1.6) {$B_{y,i,\rm MG}^{\prime\prime\prime}$};
\node at (9.8,1.) {$B_{z,i,\rm MG}^{\prime\prime\prime}$};
\node at (8.5,1.5) {${\mathbf B}_{i,\rm MG}^{\prime}:$};

\end{tikzpicture}
~\\[2mm]

The above diagram shows that the decomposition of ${\mathbf B}_{i}^{\prime}$ and ${\mathbf B}_{i,\rm MG}^{\prime}$ in the new cartesian coordinate frame $(\hat{e}_1, \hat{e}_2, \hat{e}_3)$, where $\hat{e}_1$ is parallel to the normal vector $\mathbf{\hat{A}}_{ij}$ of virtual interface between particle pair of $i$ and $j$.  Compared to ${\mathbf B}_{i}^{\prime}$ and ${\mathbf B}_{j}^{\prime}$, the modified ${\mathbf B}_{i,\rm MG}^{\prime}$ and ${\mathbf B}_{j,\rm MG}^{\prime}$ only changes the amplitude along the normal direction of the virtual surface. This is due to the fact that the increment of the reconstruction value $c_iQ_{ij}\cdot(\textbf{x}_{ij}-\textbf{x}_i) = {-}c_i|\textbf{x}_{ij}-\textbf{x}_i|^2\mathbf{\hat{A}}_{ij}$, is always parallel to the normal vector of the virtual interface. In addition, since the increment is $o(|\textbf{x}_{ij}-\textbf{x}_i|)$, in the well-designed numerical scheme, this modification also will not change the reconstruction accuracy of the magnetic field.

{
Note that in our method, the slope-limiter is only applied to the `initial' magnetic field gradient; consequently, subsequent modifications to the gradient may ``slightly re-violate'' the original slope-limiter condition. Unlike \citet{Hopkins+2016b}, we do not attempt any iterative slope correction, because we observed that exact divergence-free preservation itself also maintains stability during long-term evolution. Numerical experiments confirm that our MG approach performs well in practice despite  potential slope-limiter violation. Further exploration will be valuable for various numerical conditions in the future.
}

In this section, we introduce an approach for modifying the gradient of magnetic field. It is worth noting that the modification will only make appropriate adjustments to the reconstructed values of magnetic field and does not alter the conservatism. Thus, our approach can be extended to other MHD numerical schemes.

\section{Numerical Tests}
\label{sect:analysis}
In the following, we run a series of classic MHD problems to test the MG method,  and compare with the numerical results of {\footnotesize GIZMO} and CG methods, respectively.

\subsection{Brio-Wu shocktube}
\label{sec:num1}
We first conduct the Brio-Wu shocktube test. In the 2D periodic boundary condition (box size of $x \in (0,4)$ and $y \in(0,0.25)$), we employ $896\times56$ particles and initialize left-state vector as $(\rho, v^1, v^2, v^3, B^1, B^2, B^3, {P_{\rm gas}}) = \{1, 0, 0, 0, 0.75, 1, 0, 1\}$  and right-state $\{0.125, 0, 0, 0, 0.75, -1, 0, 0.1\}$ with $\gamma =2$, following the initial condition in \citet{Brio+1988}. The region is long enough in the x-direction to ensure that the evolution of the shock wave is not affected by the boundary during simulation run. 

\begin{figure*}[htpb]
\centering
\includegraphics[width=1.\linewidth]{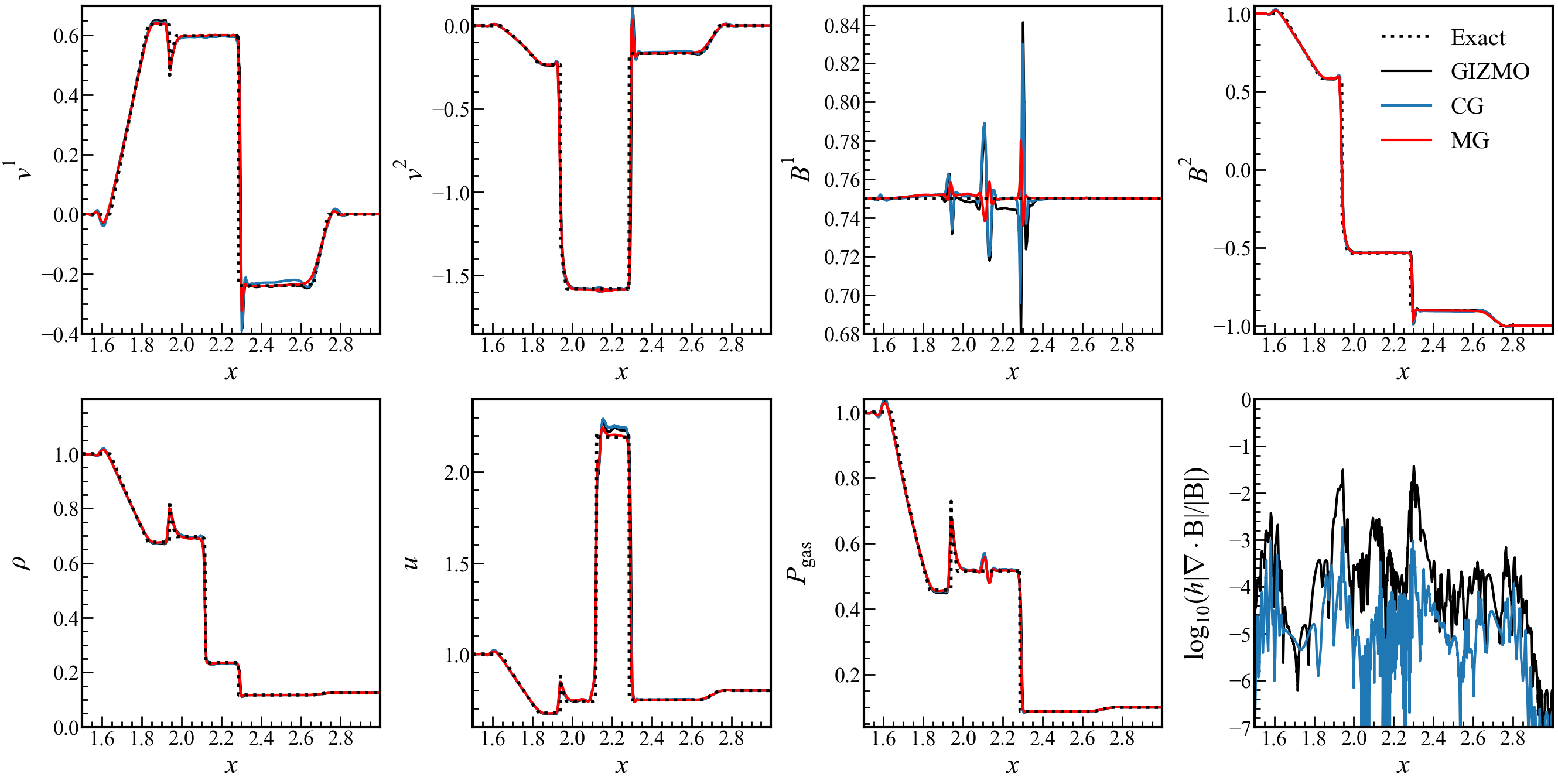}
\caption{{ Brio-Wu shock tube. We present that density, velocity, magnetic field, internal energy, pressure and the map of the divergence error $\text{log}_{10}(h_i|\nabla\cdot \mathbf{B}_i|/|\mathbf{B}_i|)$ are evolving by {\footnotesize GIZMO} method (dark line), CG method (blue line), MG method (red line)  and exact solution (dashed line)  at $t=0.2$, respectively.}}
\label{Fig1}
\end{figure*}

This numerical experiment is designed to measure the ability to capture MHD shocks,  rarefaction waves and contact discontinuities. This test is challenging for meshless methods, especially in the evolution of the first component of the magnetic field. We present the results of the MG method, and compare it with the CG method, and {\footnotesize GIZMO} scheme. Note that {the CG and {\footnotesize GIZMO} methods employ the Powell and Dedner cleaning schemes, while the MG method does not whatsoever}. Fig.~\ref{Fig1} shows the numerical results of density, velocity, magnetic field, internal energy, pressure and divergence error for MG, CG, and {\footnotesize GIZMO} methods at time $t = 0.2$, respectively.

Our MG method maintains the divergence-free constraint $\nabla \cdot \mathbf{B}=0$ to machine precision so that the amplitude curve is not been observed in the {right-bottom} panel. On contrast,   both {\footnotesize GIZMO} and CG exhibit divergence errors throughout the entire evolution region, in particular, reaching their peak in the shock wave discontinuity position.  In terms of the evolution of the first component of the magnetic field, both {\footnotesize GIZMO} and CG methods exhibit wildly oscillations, but our MG method effectively reduces the amplitude of this oscillation.

\subsection{Advection of a field loop}
\label{sec:num2}
Next, we move to the test of advection of a field loop. This is an essential test to verify for vanishing the divergence of magnetic field in MHD simulation. Meanwhile, it is also sensitive to the dissipation degree of numerical methods. The advection of field loop is initialized in a periodic squared domain, we set initial parameters, inside a circle of $R=\sqrt{x^2+y^2}<0.3$ about the geometric center, we set $(\rho, B^1,B^2)=(2,B_0y/R,B_0x/R)$ with $B_0=10^{-3}$. At outer region, we set $(\rho, B^1,B^2)=(1,0,0)$. In the full squared domain, we set $({P_{\rm gas}},v^1,v^2)=(1,2,0.5)$, $\gamma = 5/3$. 

\begin{figure}[htbp] 
\centering
\includegraphics[width=1.0\linewidth]{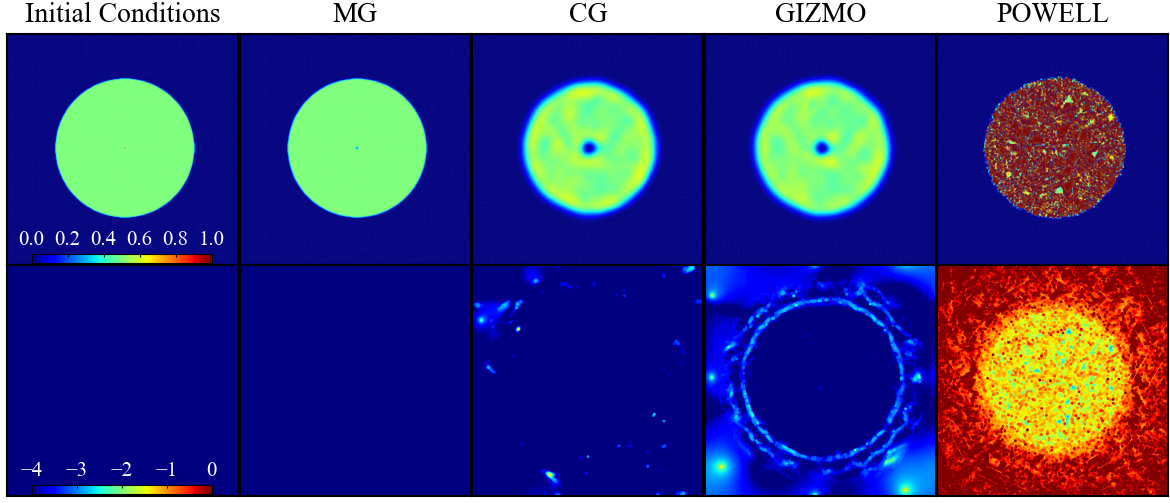}
\caption{{Advection of a field loop. The first column is the initial state, other four columns show the results at $t = 20$, we compare four methods: MG, CG, {\footnotesize GIZMO}, POWELL as labelled (left-to-right). The ideal numerical result should maintain a perfect circle, say, scale and amplitude, keep the same as the initial state. Each column shows a map of magnetic pressure ${\mathbf B}_i^2/2*10^6$ (up), and the map of the divergence error $\text{log}_{10}(h_i|\nabla\cdot \mathbf{B}_i|/|\mathbf{B}_i|)$ (bottom), with values following the color bar. All test runs employ $256^2$ particles. 
 }}
\label{Fig2}
\end{figure}

We present the results of our MG method, and compare it with the CG method, {\footnotesize GIZMO} scheme and POWELL scheme.
Fig. \ref{Fig2} shows the numerical results of a map of magnetic pressure $|{\mathbf B}_i|^2/2 \times 10^6$ and divergence error $\text{log}_{10}(h_i|\nabla \cdot {\mathbf B}|_i/|{\mathbf B}|_i)$ for MG, CG, {\footnotesize GIZMO} and Powell methods at time $t = 20$. Our MG method performs incredibly well. The magnetic pressure almost not decays over evolution, but remains constant. It comes from magnetic field divergence error to machine accuracy in MG method. In comparison, the CG method and {\footnotesize GIZMO} both show a large dissipation and the magnetic pressure decayed or amplified with time. This is mainly caused by a certain magnitude of magnetic field divergence error. The magnetic pressure of Powell scheme shows a sharp increase and oscillation, mainly due to the lack of the Dedner cleaning term that suppresses magnetic field divergence errors, resulting in a significant increase in magnetic field divergence.

\subsection{2D Orszag-Tang vortex}
\label{sec:num3}
Next, we conduct the 2D Orszag-Tang vortex problem. The 2-D compressible Orszag-Tang vortex \cite{Orszag+1979} is also initialized in a periodic squared area.  We set initial parameters to $\gamma = 5/3$, $\rho = {25}/{(36\pi)}$, ${P_{\rm gas}} = {5}/{(12\pi)}$, $(v^1,v^2) = \{ -\sin(2\pi y), \sin(2\pi x) \}$, and $(B^1,B^2) = \{ -\sin(2\pi y)/\sqrt{4\pi}, \sin(4\pi x)/\sqrt{4\pi} \}$. 

\begin{figure}[htbp] 
\centering
\includegraphics[width=1.0\linewidth]{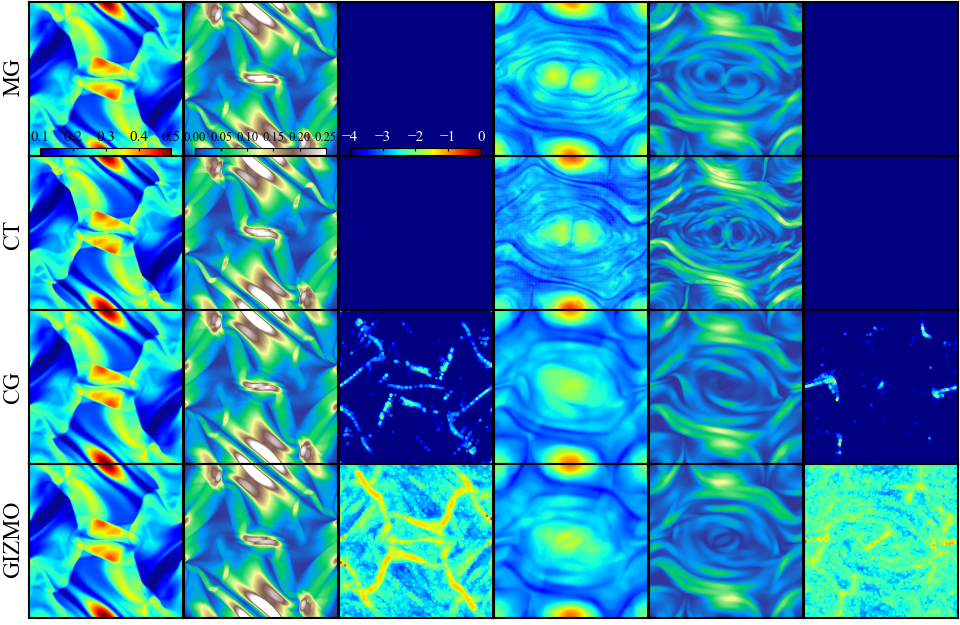}
\caption{{2D Orszag-Tang vortex. For each column, we compare four methods: MG, CT, CG, {\footnotesize GIZMO}, as labelled (top-to-bottom). Each three columns shows the density (left), magnetic pressure (middle) and the map of the divergence error $\text{log}_{10}(h_i|\nabla\cdot \mathbf{B}_i|/|\mathbf{B}_i|)$ (right), with values following the colorbar. {All meshless runs contain $256^2$ particles and CT employs a $256^2$ grid.} The results in left three columns at $t = 0.5$ and right three columns at $t = 4.0$. 
 }}
\label{Fig3}
\end{figure}

\begin{figure}[htbp] 
\centering
\includegraphics[width=1.0\linewidth]{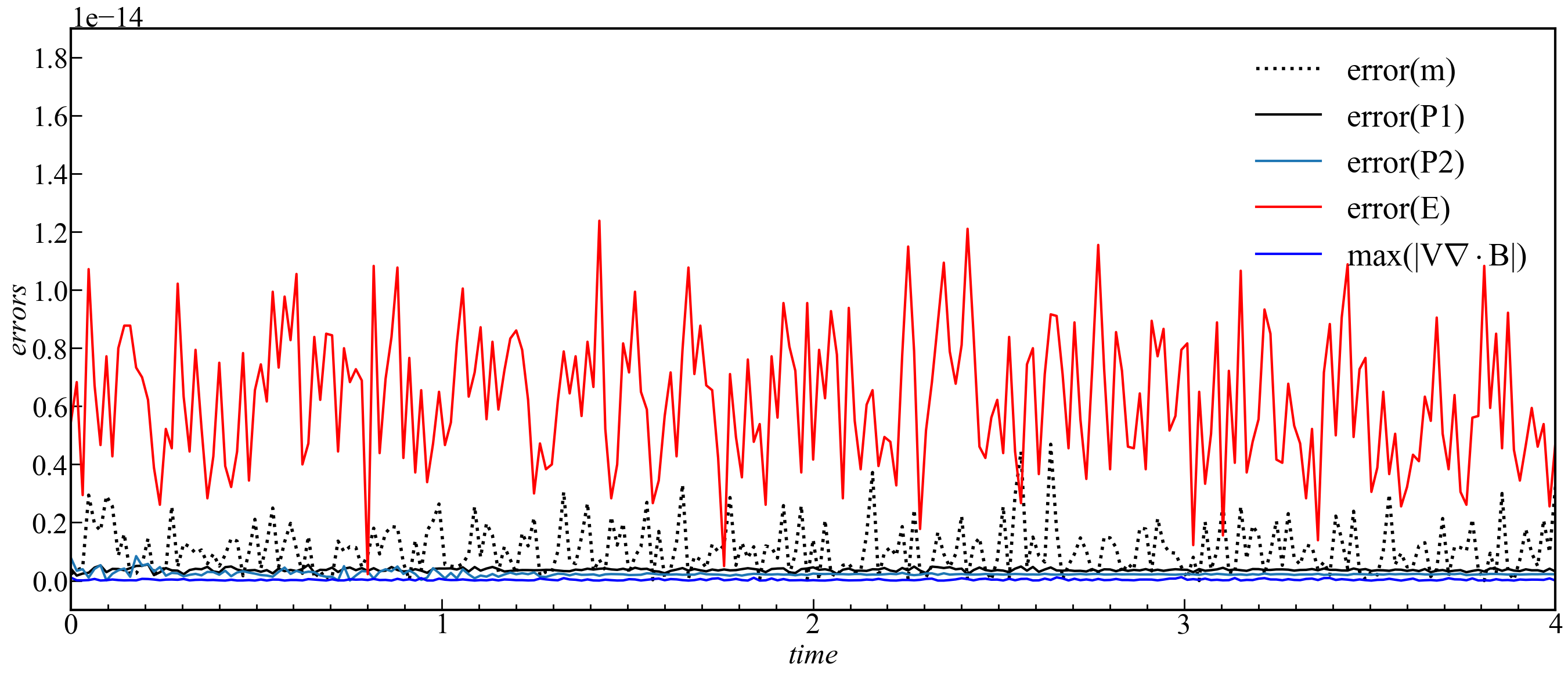}
\caption{{2D Orszag-Tang vortex. We presents the conservation results of the MG method that total mass $m$ error, total momentum (two components $P1,P2$) error, total energy $E$ error, and the maximum magnetic flux $\max_i\{(V_i\nabla\cdot \mathbf{B}_i)\}$ for 2D Orszag-Tang vortex evolution over time [0,4]. 
 }}
\label{Fig4}
\end{figure}

\begin{figure}[htbp]
\centering
  \includegraphics[width=1.0\linewidth]{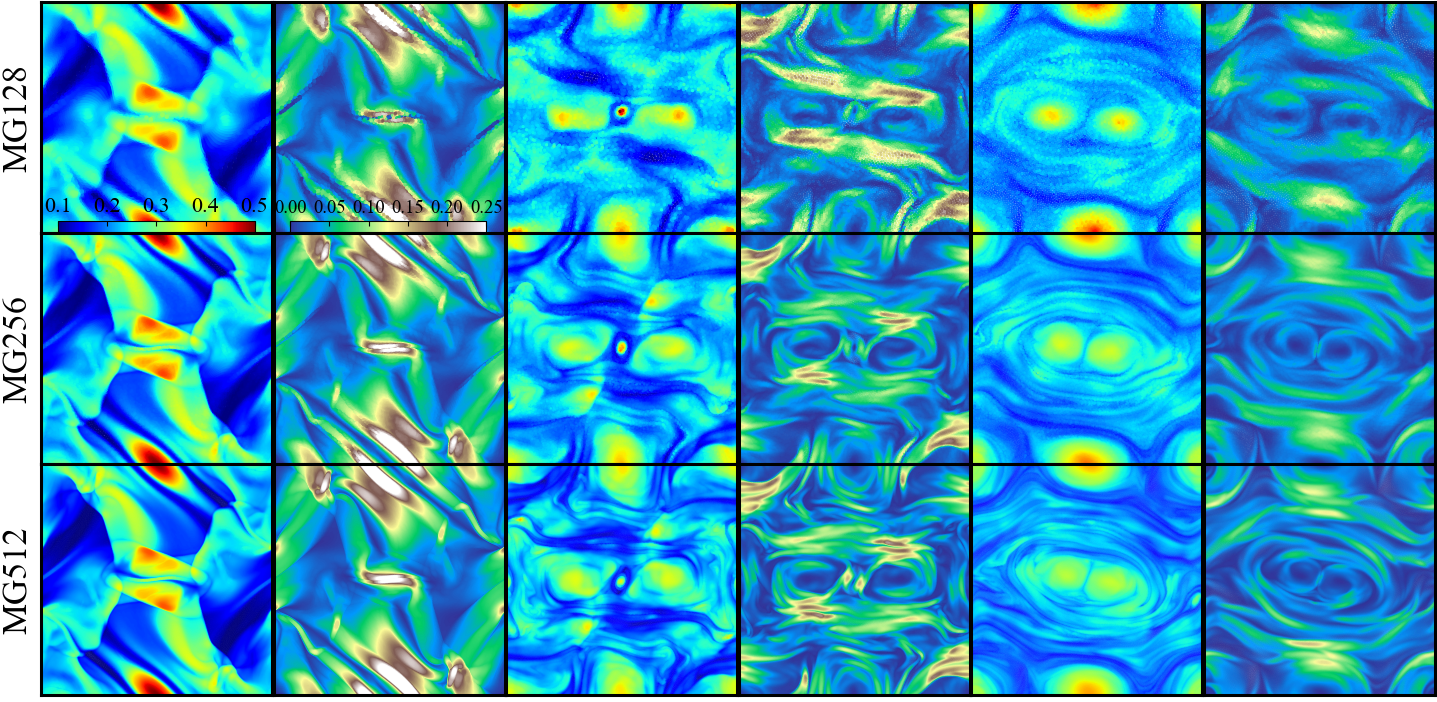}
  \caption{{\footnotesize 2D Orszag-Tang vortex. We compare the MG method at three resolutions: MG128 ( $128^2$ particles), MG256 ( $256^2$ particles), and MG512 ( $512^2$ particles), from top to bottom. Each two column present density (left) and magnetic pressure (right), respectively. The results are shown at $t=0.5$  (left), $t=2.0$  (middle), and $t=4.0$  (right).}
  }
  \label{Fig-r3}
\end{figure}

\begin{figure}[htbp]
\centering
  \includegraphics[width=1.0\linewidth]{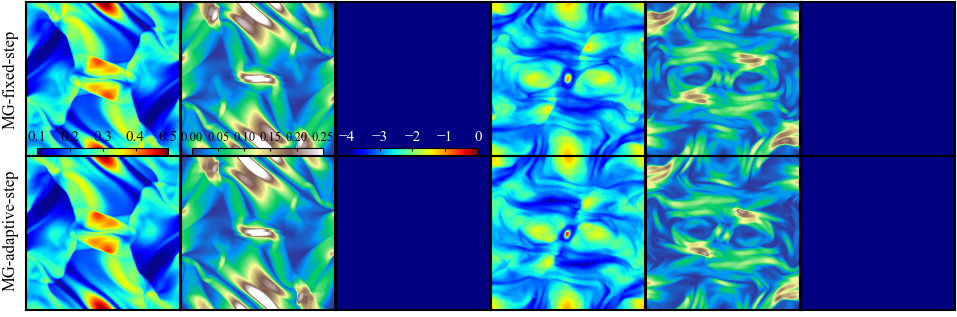}
  \caption{{\footnotesize 2D Orszag-Tang vortex. For each column, we compare two methods: MG with fixed time step, MG with adaptive time step, as labelled (top-to-bottom). Each three columns shows the density (left), magnetic pressure (middle) and the map of the divergence error $\text{log}_{10}(h_i|\nabla\cdot \mathbf{B}_i|/|\mathbf{B}_i|)$ (right), with values following the colorbar. The divergence error is measured at coarse step synchronization points.  {All runs contain $256^2$ particles.} The results in left three columns at $t = 0.5$ and right three columns at $t = 2.0$. }}
  \label{Fig-r1}
\end{figure}

We present the results of the MG, { CT,} CG and {\footnotesize GIZMO} scheme. { The numerical results of the CT method are calculated by {\footnotesize ATHENA} with a third-order precision scheme~\citep{athena2020}. }
Patterns of the density and magnetic pressure are shown in Fig.~ \ref{Fig3}. All of methods produce qualitatively similar results (except for the divergence errors) at the time of $t = 0.5$ (left three columns). However, after $t = 4.0$ (right three columns), significant difference is observed in both of density and magnetic pressure. The {\footnotesize GIZMO} method presents highly dissipative results. Although the CG method reduces the magnetic field divergence errors by several orders of magnitude. It works well  in a short-term evolution but fails in long-term evolution.

Our method performs better results of density and magnetic pressure, which attributed to its machine accuracy of divergence-free constraint and lower dissipation. At $t = 0.5$ and $t = 4.0$, a certain degree of divergence error was observed in the CG and {\footnotesize GIZMO} methods, which {causes} their numerical dissipation. { Meanwhile, the MG results match the central vortex pattern of high-order CT method at $t = 4.0$.}

{Fig. \ref{Fig4} present the error level of the conservations in the MG method. In this Figure,} The label of `error$(\rm m)$' denotes the absolute value of the total mass deviation with respect to the initial one during the evolution process. The similar definition as to the total momentum (two components $P1,P2$) and total energy $E$. We also present the evolution of the maximum absolute value of magnetic flux $V\nabla\cdot \mathbf{B}$ among all particles. These errors all achieve machine precision, indicating that our MG method provides an accurate divergence-free constraint of the magnetic field.

{ 
To verify the robustness of the MG method and its dependence on the resolution, especially for long-term evolution, we carried out simulations with different resolutions of $128^2$, $256^2$ and $512^2$ particles (labeled as MG128, MG256 and MG512, respectively) for the same initial condition. Fig. \ref{Fig-r3} presents a comparison of numerical results for the three resolutions, which appear highly consistent. For instance, a two-blob pattern exists at the center of all simulations $t = 4.0$.
}

{
However, the MG method is computationally expensive due to the numerical solution of sparse linear equations. To accelerate it, we attempt to use MG correction only for coarse step synchronization, with CG sub-steps in an adaptive-step scheme. We examine whether this approach induces instability or significantly alters the results. We perform fixed-step and adaptive-step Orszag-Tang vortex simulations until $t=2.0$, keeping all other conditions identical. Fig. \ref{Fig-r1} presents that the MG-adaptive-step and MG-fixed-step methods generate similar numerical results.
}

\subsection{Magnetized blast wave and magnetic rotor}
\label{sec:num4}
Now, we consider a strong magnetized blast wave and a magnetic rotor test in a 2D periodic unit size . For the strong magnetized blast wave, we take initial conditions following  \citet{Hopkins+2016a}: $\gamma=5/3$, density $\rho = 1$, velocity $(v^1,v^2) = \{ 0, 0 \}$ and $(B^1,B^2) = \{ 1/\sqrt{2}, 1/\sqrt{2} \}$. The center of the domain as the origin, pressure ${P_{\rm gas}}=10$ for $r<0.1$, $r=\sqrt{x^2+y^2}$, otherwise ${P_{\rm gas}}=0.1$. For the magnetic rotor test, we initialize a 2D domain with $\gamma = 7/5$, pressure ${P_{\rm gas}}=1$, $(B^1,B^2)=\{5/\sqrt{4\pi},0\}$. The  origin is set at the center of the domain and there are density $\rho=10$ and velocity $(v^1,v^2)=\{-2y/r_0,2x/r_0\}$ for $r<r_0=0.1$; $\rho=1+9f(r)$ and $(v^1,v^2)=\{-2yf(r)/r,2xf(r)/r\}$ with $f(r)=(r_1-r)/(r_1-r_0)$ for $r_0<r<r_1=0.115$, otherwise $\rho=1$ and $(v^1,v^2)=\{0,0\}$, in the region of  $r=\sqrt{x^2+y^2}$. MHD rotor was first introduced by \citet{Balsara+1999}, and has become a standard test for 2D MHD problem. 

For both tests, we also present the comparison of the MG method with the CG method and {\footnotesize GIZMO} scheme.
Fig. \ref{Fig5} shows the numerical results of density, gas pressure, magnetic pressure and the divergence error of magnetic field for MG, CG and {\footnotesize GIZMO} methods. Our MG method achieved the $\nabla \cdot \mathbf{B}=0$ constraint to machine precision, but CG and {\footnotesize GIZMO} methods retain the divergence errors. All the methods produce qualitatively similar results for the fluid variables (except for the divergence errors). These two numerical experiments aim at extreme MHD problems.

\begin{figure}[htbp] 
\centering
\includegraphics[width=1.0\linewidth]{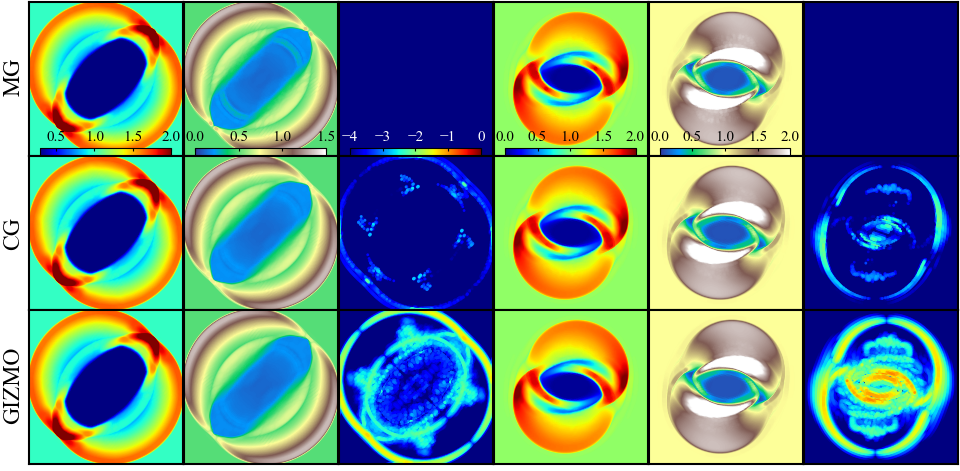}
\caption{{MHD blast wave at time $t=0.2$ (left three columns) and MHD rotor at time $t = 0.15$ (right three columns). For each column, we compare three methods: MG, CG, {\footnotesize GIZMO}, as labelled (top-to-bottom). Each three columns show a map of a fluid quantity (left), magnetic pressure (middle) and the map of the divergence error $\text{log}_{10}(h_i|\nabla\cdot \mathbf{B}_i|/|\mathbf{B}_i|)$ (right), with values following the color bar. All runs here have $256^2$ particles.
Left: MHD blast wave, showing density (the first column). 
Right: MHD rotor, showing gas pressure (the fourth column). 
  }}
\label{Fig5}
\end{figure}

\subsection{Magnetorotational instability}
\label{sec:num5}
In the subsection, we test Magnetorotational Instability. Magnetorotational Instability (MRI) is an essential numerical experiment in astrophysics \citep{Balbus+1991,Balbus+1998,Guan+2008}. MRI is carried out in  a axisymmetric 2D shearing box, $-0.5<x<0.5$ and $-0.5<z<0.5$. Following \citet{Hopkins+2016a}, we setup the density field $\rho=1$, pressure ${P_{\rm gas}}=1$, velocity $(v^1,v^2,v^3)=\{\delta v,-q x, \delta v\}$ and magnetic field $(B^1,B^2,B^3)=\{0,0,B_{0}\sin(2\pi x)\}$, where $ q =3/2$ and  $B_0=\sqrt{15}/(8\pi m)$ with the mode number $m=4$. The velocity $\delta v \in[-0.005,0.005]$ is randomly generated. In this test, it is necessary to add the momentum equations with an additional source terms, $D(\rho \mathbf{v})/Dt = -2(\Omega\hat{z})\times(\rho \mathbf{v}) + 2\rho q \Omega^2x\hat{x}$, which correspond to the Coriolis and the centrifugal forces, respectively. All variables undergoes a periodic boundary condition, except for $v^2$ in $x$ axis. In specific, $v^2(x,z)=v^2(x+ n_x L_x,z+ n_z L_z) + n_x q \Omega L_x$, the angular speed $\Omega = 1$. 

\begin{figure}[htbp] 
\centering
\includegraphics[width=1.0\linewidth]{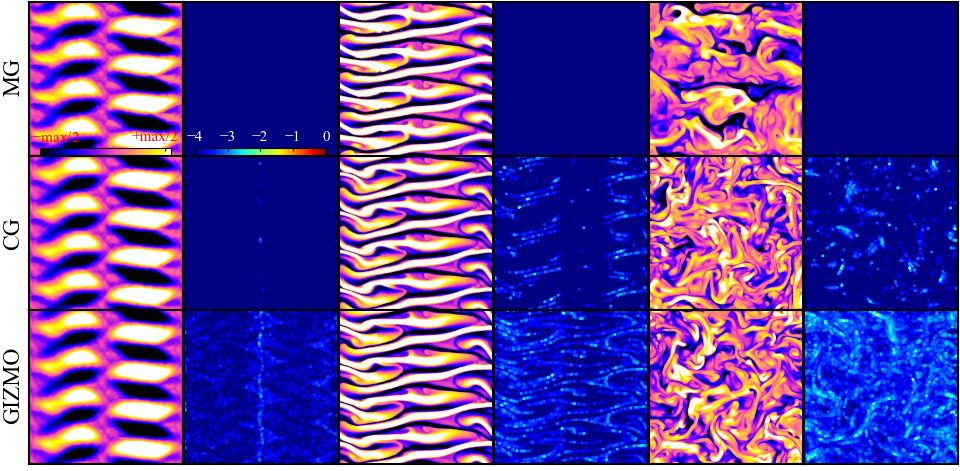}
\caption{{Magnetorotational instability (MRI). For each column, we compare three methods: MG, CG, {\footnotesize GIZMO}, as labelled (top-to-bottom). Each pair of columns shows the azimuthal/toroidal component $B_y$ of the magnetic field (left), and the map of the divergence error $\text{log}_{10}(h_i|\nabla\cdot \mathbf{B}_i|/|\mathbf{B}_i|)$ (right), with values following the colorbar, `max' represents its maximum absolute value. All runs here have $256^2$ particles.
Left: showing the results at $t = 12$. 
Middle: showing the results at $t = 16$. 
Right: showing the results at $t = 19$. 
}}
\label{Fig6}
\end{figure}

MRI is a special numerical experiment, where the entire evolution is carried out on a two-dimensional unit plane, but the velocity and magnetic field are three-dimensional  variables. We present the results of the MG method, and compare it with the CG and {\footnotesize GIZMO} scheme at $t = 12$, $t = 16$ and $t = 19$. The azimuthal/toroidal component $B_y$ of magnetic field gradually goes up, and then pass through the axis ($x = 0$) and boundary in pairs of positive and negative values through advection. Gradually turbulence happens. In Fig.~\ref{Fig6}, {it} is obvious that our MG method preserves the magnetic field divergence-free constraint, but other methods fail. The divergence errors of the {\footnotesize GIZMO} and CG methods continue to increase as the evolution. The results of CG and {\footnotesize GIZMO} methods seem to have obstacles of the magnetic field component $B_y$ passing to the opposite side in pairs. Compared to the {\footnotesize GIZMO} and CG methods, our MG method better captures advection.

\subsection{3D Orszag-Tang vortex}
\label{sec:num6}

Finally, we attempted to execute a  3D simulation using our MG method to verify whether it remains effective in preserving the divergence-free constraint $\nabla\cdot \mathbf{B}_i = 0$ in 3D MHD problems-- the 3D Orszag \& Tang vortex. We follow the work of \citet{Helzel+2011} to set up the initial condition. In a periodic 3D unit box, the initial variables are set as $\rho = {25}/{(36\pi)}$, ${P_{\rm gas}} ={5}/{(12\pi)}$, 
\begin{equation}\notag
(v^1,v^2,v^3) = \{-[1+\epsilon_p\sin(2\pi z)]\sin(2\pi y),  [1+\epsilon_p\sin(2\pi z)]\sin(2\pi x), \epsilon_p\sin(2\pi z) \}
\end{equation}
and
\begin{equation}\notag
(B^1,B^2,B^3) = \left\{ -\frac{\sin(2\pi y)}{\sqrt{4\pi}}, \frac{\sin(4\pi x)}{\sqrt{4\pi}}, 0 \right\}.
\end{equation}

We choose $\epsilon_{p} = 0.2$. The adiabatic index of gas is $\gamma = 5/3$. In this 3D case, we carry out the simulation with $64^3$ particles evolves  to the time  of $t = 2.5$. We found our method still perform well up to  machine precision for  $\nabla \cdot \mathbf{B}=0$. 

Because of  periodic condition, we show the fluid  variables at three surfaces $x=0$, $y=0$ and $z=0$ in Fig. \ref{Fig7}-\ref{Fig9}. We confirm that MG method can effectively preserves the $\nabla \cdot \mathbf{B}=0$ constraint to machine precision, even in 3D problems.

\begin{figure}[htbp] 
\centering
\subfigure{
\includegraphics[height=0.28\textheight]{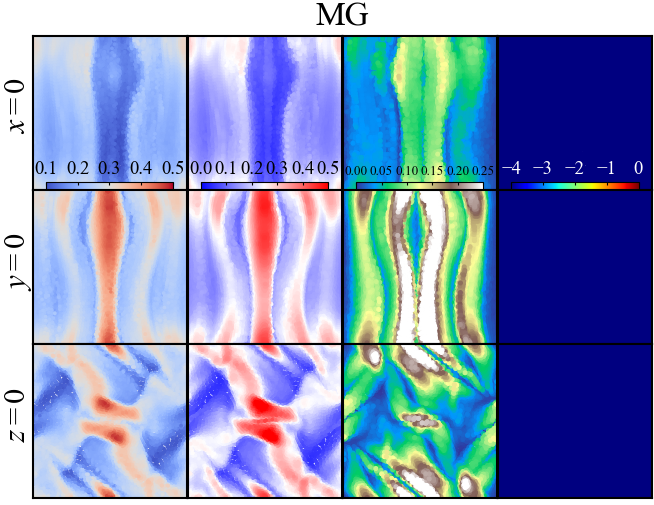}
}
\subfigure{
\includegraphics[height=0.28\textheight]{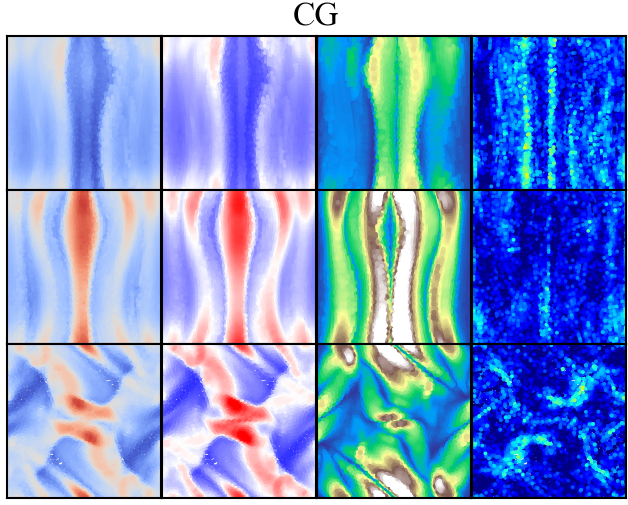}
}
\caption{{3D Orszag-Tang vortex at $t = 0.5$. We compare two methods: MG, CG, as labelled (left-to-right). For each row, we show the fluid quantitys at three surfaces $x=0$, $y=0$ and $z=0$ as labelled (top-to-bottom). Each four columns show the density (left), gas pressure (middle left), magnetic pressure (middle right) and the map of divergence error $\text{log}_{10}(h_i|\nabla\cdot \mathbf{B}_i|/|\mathbf{B}_i|)$ (right) for MG method (left four columns) and CG method (right four columns), with values following the colorbar.
 }}
\label{Fig7}
\end{figure}

\begin{figure}[htbp] 
\centering
\subfigure{
\includegraphics[height=0.28\textheight]{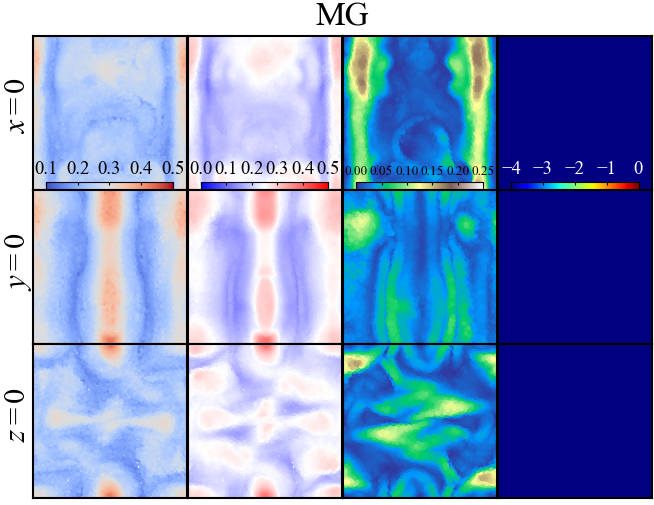}
}
\subfigure{
\includegraphics[height=0.28\textheight]{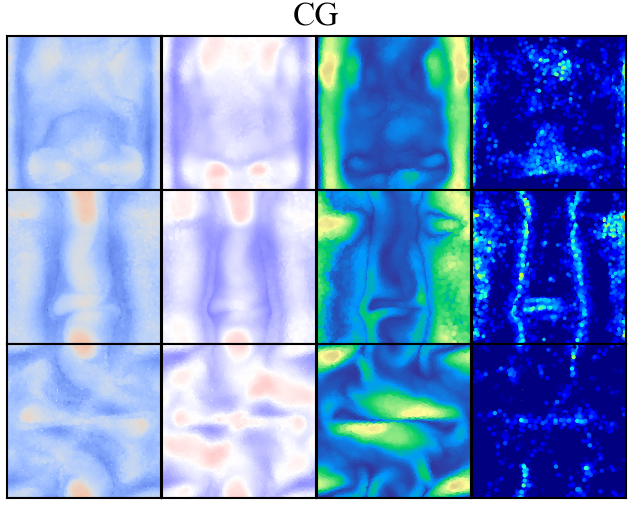}
}
\caption{{3D Orszag-Tang vortex at time $t = 2.0$. We compare two methods: MG, CG, as labelled (left-to-right). The position and fluid quantity represented by each pattern are the same as that of Fig. \ref{Fig7}.
 }}
\label{Fig8}
\end{figure}

\begin{figure}[htbp] 
\centering
\subfigure{
\includegraphics[height=0.28\textheight]{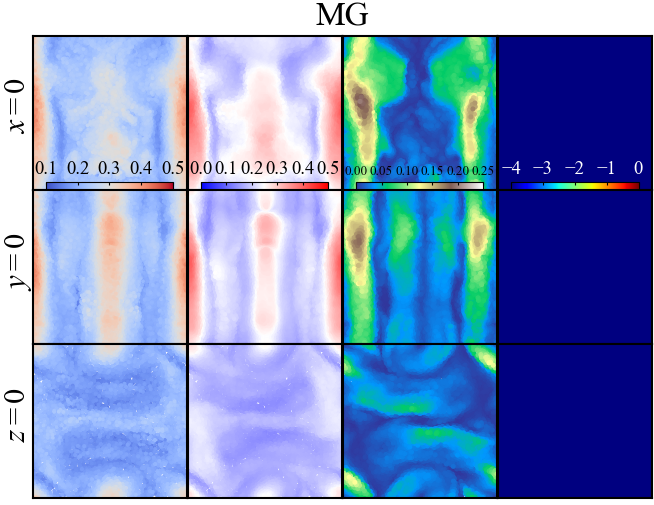}
}
\subfigure{
\includegraphics[height=0.28\textheight]{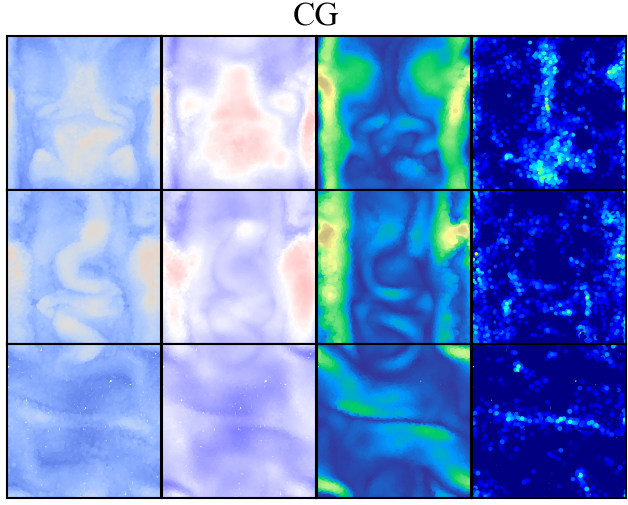}
}
\caption{{3D Orszag-Tang vortex at time $t = 2.5$. We compare two methods: MG, CG, as labelled (left-to-right). The position and fluid quantity represented by each pattern are the same as that of Fig. \ref{Fig7}.
 }}
\label{Fig9}
\end{figure}

{Additionally, we present the results of the MG , CG scheme. Patterns of the density, gas pressure, magnetic pressure, and divergence error are shown in Fig. \ref{Fig7}-\ref{Fig9}. Both methods produce qualitatively similar results at $t = 0.5$ (see Fig. \ref{Fig7}), except for divergence errors. Fig. \ref{Fig8} and \ref{Fig9} indicate that the MG method produces more robust dynamic evolution profiles compared to the CG method, particularly at the surface $y = 0$. This robustness stems from the MG method's maintenance of the divergence-free constraint. Conversely, the CG method exhibits significant advective errors starting from $t = 2.0$, verified during the advection test in the field loop. At $t = 2.5$, the profiles given by the CG method shows pronounced deformation and severe dissipation at the surface $y = 0$, alongside asymmetric distribution patterns at the surface $z = 0$.

\section{Conclusion}
\label{sect:conclusion}

{
In this work, we present a Lagrangian  meshless modified-gradient method with a Godunov-type magnetohydrodynamic solver. The MG method could be considered as an extension and modification with respect to  constrained-gradient (CG) methods \cite{Hopkins+2016b}. Its regularization on reconstruction of magnetic field in the virtual interface, ensures the divergence-free condition during simulation run so that our method remains a conservative scheme.

According to results of numerical experiments, we confirm that our MG method can maintain the global divergence-free constraint $\nabla\cdot \mathbf{B}_i = 0$ to machine precision. 
In the Brio-Wu shock tube test, our MG method strongly decrease the oscillation of the first component of magnetic field, which is difficult in many meshless methods. The MG method almost eliminates the advection errors in the example of advection of a field loop test. It fully confirms the advantage of our method on the divergence-free constraint. In the 2D Orszag-Tang vortex test, our method performance better than CG code on divergence errors as well. 
Due to the exact divergence-free constraint, the effective work from Lorentz force perfectly vanishes in our method. Thus, the dissipation of fluid quantities, e.g. density and magnetic field, are also reduced. On the contrast, such effect is not observed a long-term evolution with CG method. 
In the cases of the magnetized blast wave, magnetic rotor, and magnetorotational instability, the MG method also works well for those special extreme problems.

In conclusion, the novel MG method is robust and competitive to constrain magnetic divergence in meshless numerical MHD simulations. We do not claim that the divergence issue has been solved by the new method. Because our MG method requires solving a large system of symmetric sparse linear equations, which will bring new challenge to the efficiency of numerical method. It expects to develop more parallel techniques for massive scale simulations. 

}

\section{Acknowledgements}
We thank the anonymous referee for their thorough review and recommendations. We acknowledge the support from the National Key Research and Development Program of China (Grant No. 2023YFB3002501), the National Natural Science Foundation of China (No. 12301504), the Natural Science Foundation of Hubei Province (No. 2024AFB722), Hubei Province University's outstanding young and middle-aged scientific and technological innovation team project (No. T2023020), Talent Introduction Project of Hubei Normal University in 2023 (Grant No. HS2023RC051). YT acknowledge the support from NSFC grant (No. 12171466, 92470119).


\end{document}